\documentclass[12pt,preprint]{emulateapj}

\def\hst{{\it HST}}

\def\swift{{\it Swift}}
\def\hete{{\it HETE-2}}

\def\H0{{\rm ~km~s^{-1}~Mpc^{-1}}}

\newcommand{\plusminus}[2]{\ensuremath{^{+#1}_{-#2}}}
\newcommand{\nh}{\ensuremath{\mathrm{N}_\mathrm{H}}}

\shorttitle{The afterglow and host of GRB 060121}
\shortauthors{A.J. Levan et al.}

\begin{document}


\title{The faint afterglow and host galaxy of the short-hard GRB 060121}


\author{A.J. Levan\altaffilmark{1,2},
 N. R. Tanvir\altaffilmark{1},   A.S. Fruchter\altaffilmark{3}, E.Rol\altaffilmark{4},
 J.P.U. Fynbo\altaffilmark{5}, J. Hjorth\altaffilmark{5},  G. Williams\altaffilmark{6},\\
E. Bergeron\altaffilmark{3}, D. Bersier\altaffilmark{7}, M. Bremer\altaffilmark{8},
T. Grav\altaffilmark{9}, 
P. Jakobsson\altaffilmark{5}, K. Nilsson\altaffilmark{10}, E. Olszewski\altaffilmark{6},\\
R.S. Priddey\altaffilmark{1}, 
D. Rafferty\altaffilmark{11}, J. Rhoads\altaffilmark{12}
}

\altaffiltext{1}{Centre for Astrophysics Research, University of
  Hertfordshire, Hatfield, AL10 9AB, UK}
\altaffiltext{2}{Kavli Institute for Theoretical Physics, University of California at Santa Barbara, CA 93106,
USA}
\altaffiltext{3}{Space Telescope Science Institute, 3700 San Martin Drive, Baltimore, MD 21218}
\altaffiltext{4}{Department of Physics and Astronomy, University of Leicester, University Road,
Leicester, LE1 7RH, UK}
\altaffiltext{5}{Dark Cosmology Centre, Niels Bohr Institute, University of Copenhagen,
Juliane Maries Vej 30, 2100 Copenhagen, Denmark}
\altaffiltext{6}{Steward Observatory, University of Arizona, 933 N. Cherry Avenue, Tucson, 
AZ 85721}
\altaffiltext{7}{Astrophysics Research Institute, Liverpool John Moores University, Twelve Quays 
House, Egerton Wharf, Birkenhead, CH41 1LD, UK}
\altaffiltext{8}{Department of Physics, Bristol University, H.H. Wills Laboratory, Tyndall Avenue, Bristol BS8 1TL}
\altaffiltext{9}{Institute for Astronomy, 2680 Woodlawn Drive, Honolulu,  Hawaii 96822, USA}
\altaffiltext{10}{Tuorla Observatory, V\"ais\"al\"antie 20, FIN-21500, Piikki\"o, Finland}
\altaffiltext{11}{Department of Physics and Astronomy,
Ohio University , Athens, OH 45701, USA}
\altaffiltext{12}{Department of Physics and Astronomy, Arizona State University, P.O. Box 871504,
Tempe, Arizona, 85287, USA}

\email{levan@star.herts.ac.uk}

\begin{abstract}
We present optical and X-ray observations of the afterglow and host
galaxy of the short-hard GRB 060121. The faint R-band afterglow is
seen to decline as $t^{-0.66 \pm 0.09}$ while the X-ray falls as
$t^{-1.18 \pm 0.04}$, indicating the presence of the cooling break
between the two frequencies. However, the R-band afterglow is very
faint compared to the predicted extrapolation of the X-ray afterglow
to the optical regime (specifically, $\beta_{\mathrm{OX}} \sim 0.2$), while the K-band
is consistent with this extrapolation ($\beta_{\mathrm{KX}} \sim 0.6$), demonstrating
suppression of the optical flux. Late time {\it HST}
observations place stringent limits
on the afterglow R-band flux implying a break in the R-band
lightcurve. They also show
that the burst occurred at the edge of a faint red galaxy which most
likely lies at a significantly higher redshift than the previous
optically identified short-duration bursts.  Several neighboring
galaxies also have very red colors that are similarly suggestive of higher
redshift.  We consider possible explanations for the faintness and
color of the burst. Our preferred model is that the burst occurred at
moderately high redshift and was significantly obscured; however,
it is also possible that the burst lies at
$z > 4.5$ in which case the faintness of the R-band afterglow
could be attributed to the Lyman-break. We discuss
the implications that either scenario would have for the
nature of the progenitors of short bursts.
\end{abstract}


\section{Introduction}

The nature of the distinct subset of $\gamma$-ray bursts (GRBs) with
short-durations of $<$2 s has seen rapid progress over the past year,
following the discovery of the first X-ray afterglow to GRB 050509B
(Gehrels et al. 2005; Bloom et al. 2006). A number of other afterglows
have been discovered subsequently at X-ray, optical and radio
wavelengths (e.g. Hjorth et al. 2005a; Fox et al. 2005; Berger et
al. 2005; Soderberg et al. 2006). These observations show that
typically short bursts have fainter afterglows than those of long
duration bursts, lie at lower redshift (usually, it seems, $z<1$), and
are associated with galaxies of all types, including those with no
sign of ongoing star formation (Gehrels et al. 2005; Berger et
al. 2005).  The latter fact indicates a significant delay between the
formation of the progenitor stars, and the creation of the
GRB. Indeed, in no case has any sign of supernova emission been seen
(e.g. Hjorth et al. 2005b). These characteristics contrast markedly
with those of long bursts which originate in star forming host galaxies
(e.g. Fruchter et al. 1999; Christensen et al. 2004) at higher redshift (Jakobsson et
al. 2006) and are now known to be associated with the core-collapse of
massive stars (e.g. Hjorth et al. 2003). The preferred
model for the short-duration bursts has thus become that the majority
of them come from either neutron star - neutron star (NS-NS) or
neutron star - black hole (NS-BH) mergers (e.g. Eichler et al. 1992;
Davies, Levan \& King 2005; Lee, Ramirez-Ruiz \& Granot 2005),
although other mechanisms which produce GRBs in populations of all
ages (e.g. Usov 1992; Dermer \& Atoyen 2006; Levan et al. 2006a)
remain possible given the paucity of observational constraints so far.

Here we present optical and X-ray observations of GRB 060121.
This short-hard burst was discovered by the {\it High Energy Transient
Explorer} (\hete) and is only the fourth to have
a well-studied optical afterglow.  The afterglow of GRB 060121 is
approximately a magnitude fainter than the other examples at similar
times. In previous cases a bright host galaxy was readily identified
under the optical transient;
however in the case of GRB 060121 this is not the case. In fact, deep
{\it Hubble Space Telescope (HST)} 
observations, reported below, were necessary to locate a faint,
red galaxy at the location of the optical afterglow.

\section{Observations}

GRB 060121 was detected by \hete\  on 2006 January 21, 22:24:54.5 UT
and was localised by the Soft X-ray camera (SXC)
(Prigozhin et al. 2006).  It was
identified as a short, $t_{90} < 2$ s, and spectrally hard burst with
a peak in its $\nu f_{\nu}$ spectrum at 120 $\pm$ 7 keV (Boer et
al. 2006).  
{\it Swift} performed target of opportunity
observations of GRB 060121, beginning at 2006 January 22 01:21:37
UT. Observations with the X-ray telescope (XRT) revealed 
a relatively bright X-ray afterglow (Mangano et al. 2006).

\subsection{The Optical Afterglow}

Our first observations of GRB 060121 were taken with the Nordic Optical
Telescope (NOT) using ALFOSC, starting at 2 hours post-burst. 
Subsequent observations were obtained with the 90prime imager 
(Williams et al. 2004) on the Steward Observatory 2.3-m Bok Telescope
 and at the Wisconsin Indiana Yale NOAO 
telescope (WIYN). All of these observations were
reduced in the standard fashion within IRAF. A log of the observations
is shown in Table 1.

Inspection of the position of the X-ray afterglow of GRB 060121
revealed a faint optical source in our NOT observations (also
reported by Malesani et al. 2006). This was clearly seen to fade
between the first observations and those obtained later and thus we
identified it as the afterglow of GRB 060121 (Levan et al. 2006b). Astrometry 
was refined relative to 36 objects identified by the SDSS
which fell within the field of view of our first epoch WIYN image. The
resulting location is RA= 09$^h$09$^m$51.99$^s$, Dec=
45$^{\circ}$39$^{\prime}$45.6$^{\prime\prime}$.  We performed
photometry of the afterglow using an aperture with radius equal to the
FWHM of the images. Our photometric calibration was based on the SDSS
observations of the field. We converted the SDSS magnitudes to the standard
Cousins-R band
using the transformations given by Jester et al (2005). The
resulting lightcurve is shown in Figure 1.  The temporal decay is
best fit with $\alpha =-0.66 \pm 0.09$ (where $\alpha$ is defined as
$F_{t} \propto t^{\alpha}$). However, the $\chi^2/\mathrm{DOF} = 11.27/4$ is
relatively poor for this fit, largely due to an apparent plateau/flare between
$\sim 5$ and 11 hours after the burst, although given the small 
number of data points it is not possible to characterize this 
feature in detail. Later IR observations have been reported by Hearty et al. (2006)
and are roughly simultaneous with our final WIYN observation. Given
their reported magnitude of $Ks \sim 20$ the afterglow color at this
time was R$-$K$ \approx 5$. This is very red, corresponding to a spectral
slope (defined as $F_{\nu} \propto \nu^{\beta}$) of $\beta_{\mathrm{RK}} = -2.7$.

\begin{table}
\begin{center}
\caption{Table of photometric observations of GRB 060121}
\begin{tabular}{lllllll}
\hline
Date & UT & $\Delta t $(hours) &  Telescope & band & magnitude \\
\hline
2006-01-22& 00:23:53&      1.98  & NOT&       R &      22.65 $\pm$ 0.21\\
2006-01-22& 01:00:21&      2.59 &  NOT &    I &     21.96 $\pm$ 0.30 \\
2006-01-22& 04:02:45&      5.63  & Bok 2.3m&    R &      23.79 $\pm$ 0.19\\
2006-01-22& 04:52:06&      6.48 & Bok 2.3m  &    B &       $>24.0$\\
2006-01-22& 05:40:29&      7.25 & Bok 2.3m &    R &      $23.72 \pm 0.15$ \\
2006-01-22& 09:41:39&      11.27 &Bok 2.3m &    R &      $23.44 \pm 0.25$\\
2006-01-22& 11:32:15&      13.12  &  WIYN  &    R &     23.75 $\pm$ 0.20\\
2006-01-23& 07:58:22&      33.56  &  WIYN  &    R &      24.91$\pm$ 0.16\\
\hline
\end{tabular}
\end{center}
\tablecomments{Photometry of the afterglow of GRB 060121 obtained
at the NOT, the Bok 2.3m, and the WIYN. Magnitudes are not corrected for
the small Galactic extinction ($A_R = 0.044$).}
\label{default}
\end{table}

\begin{figure}
\begin{center}
\plotone{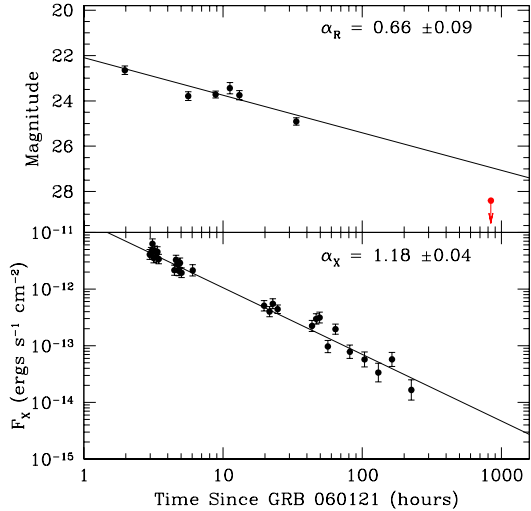}
\caption{R-band (top) and X-ray (0.2-10 keV - bottom) lightcurves
of GRB 060121, fitted with single power-law
decay indicies  as shown and discussed sections 2.1 \& 2.2.}
\label{lightcurve}
\end{center}
\end{figure}

\subsection{The X-ray afterglow}

\swift\ XRT data were reduced using the
FTOOL \texttt{xrtpipeline}. The light
curve and spectrum were extracted using \texttt{xselect}, where data
from the first 3 orbits have been filtered on grade 0 to counter any pile-up.
Later orbits use the default
filtering. A circular extraction region of radius of
71\arcsec\ was employed for the first 3 orbits where the source was
relatively bright, while a 47\arcsec\ radius circle
was used for later orbits. The background was estimated using
extraction regions around the source, and has been dynamically
subtracted for the light curve and spectrum. For both the light curve
and the spectrum, 20 counts were grouped together per bin.

The X-ray light curve (Figure 1 - lower panel) 
has been modelled with a single power law,
resulting in a fit with $\chi^2/\mathrm{DOF} = 28.52/27$ and a power
law decay index of $\alpha = -1.18 \pm 0.04$. 
The data is consistent with a monotonic evolution
out to $\sim$ 10 days since the burst, ruling out any
jet break occurring before this time. Unfortunately no
X-ray data exists for the period covering the possible optical
flare. Thus we cannot rule out a flare in the X-ray data, in
which the underlying decay was unaffected. However, since
the X-ray light curve continues as a single power law,
energy injection, as seen in GRB 051221 (Soderberg et al. 2006)
can be ruled out.

The data used for the spectrum were obtained from the first 3 orbits
where the X-ray afterglow was bright.
The spectrum can be fit with an absorbed power law, with
$\chi^2/\mathrm{DOF} = 8.11/14$. The resulting photon index is $\Gamma
= 2.33\plusminus{0.26}{0.23}$ and the absorbing column density $\nh =
1.31\plusminus{0.46}{0.24} \times 10^{21} \mathrm{cm}^{-2}$. The
estimated Galactic absorption in this direction is $1.7 \times 10^{20}
\mathrm{cm}^{-2}$ (Dickey \& Lockman 1990). Some interdependence is
seen between \nh\ and $\Gamma$, however this is insufficient to
explain the observed excess \nh.  We therefore conclude that GRB
060121 exhibits significant X-ray absorption above the Galactic
value.  This absorption estimate assumes
zero redshift. At higher redshift the rest-frame absorption may be
substantially larger.

\begin{figure}
\begin{center}
\plottwo{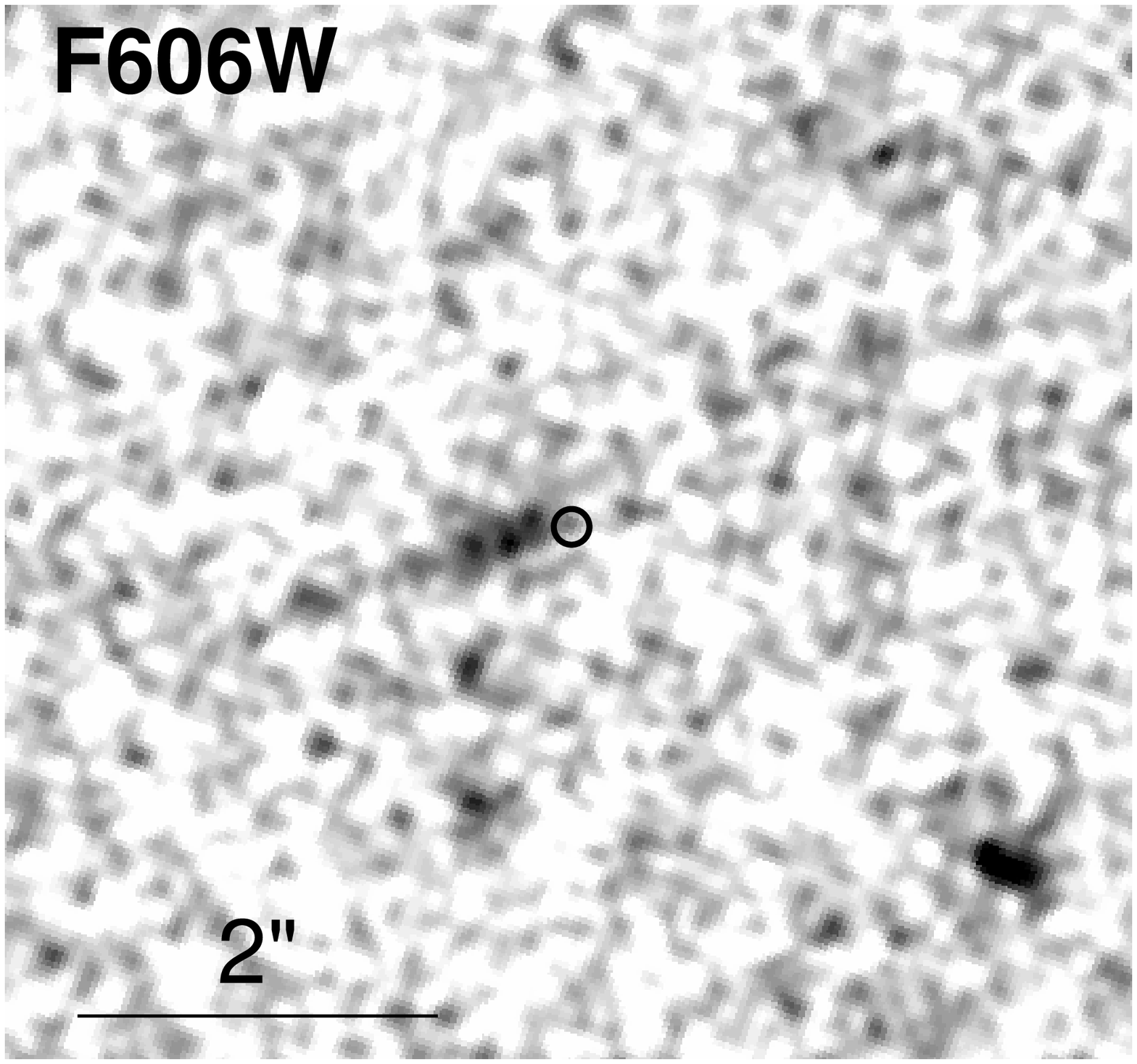}{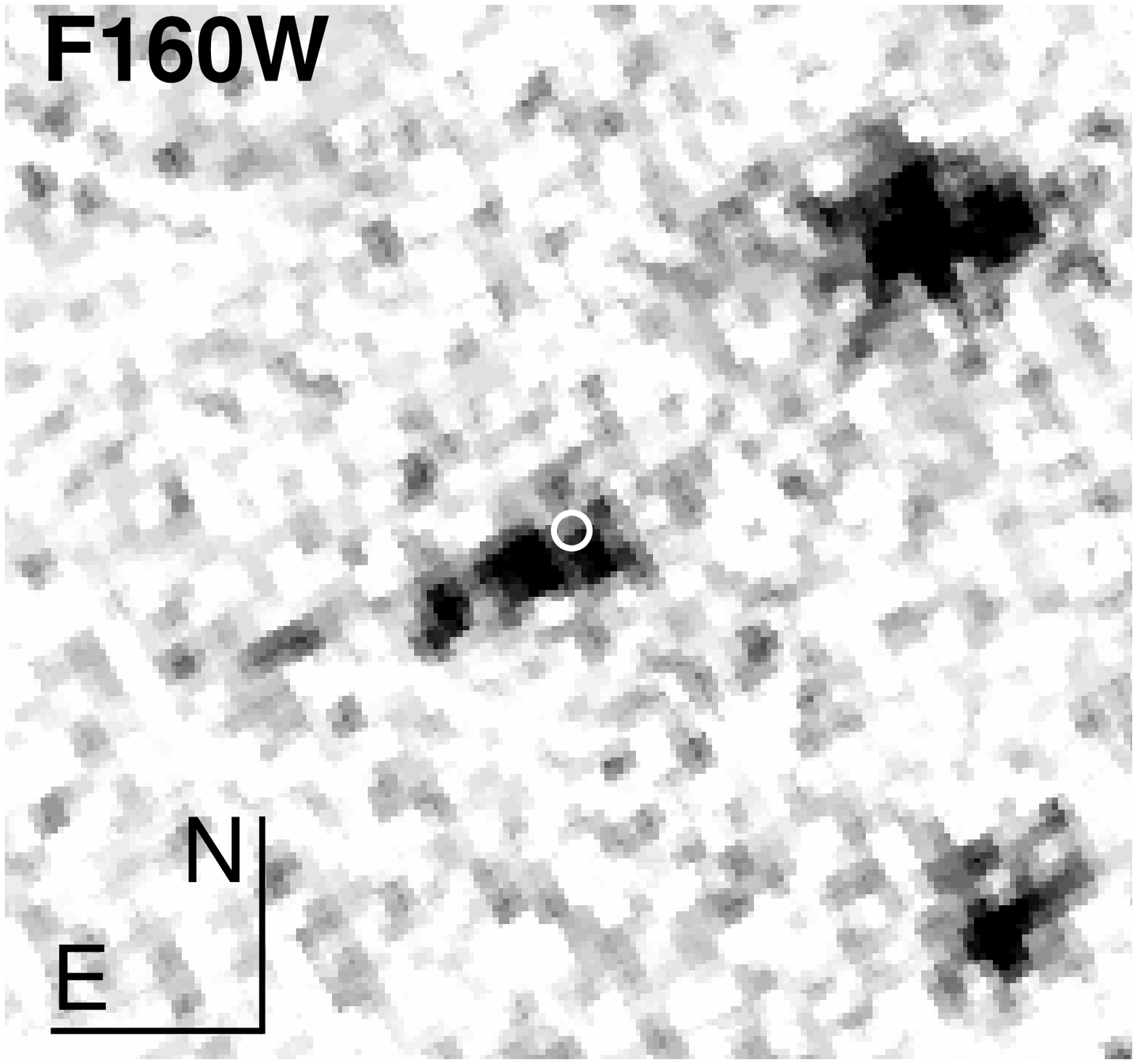}
\caption{HST observations of GRB 060121 with ACS/F606W (left - smoothed
by a gaussian of the same with as a PSF to enhance faint features) and
NICMOS/F160W (right). 
In each case the location of the burst is
marked with a white circle. As can be seen the burst lies on an IR
brighter region which may indicate either a color gradient or a
continuing contribution from the afterglow in the IR.  
An 
ERO can be seen to the Northwest of the host in NICMOS.}
\label{lightcurve}
\end{center}
\end{figure}

\subsection{HST observations}

GRB 060121 was observed with \hst\ on February 27th 2006. Observations
were obtained with both NICMOS/F160W and
ACS/WFC/F606W. A total of 4608 s of
observations were obtained using NICMOS and 4400 s using ACS.  
Astrometry was performed between the
ground based observations of GRB 060121 and the ACS observations, and
subsequently the NICMOS observations. Using 10 point sources in common
to each image we were able to obtain a registration of the afterglow
position on the ACS images accurate to 0.1\arcsec ($1 \sigma$).
Inspection of this location reveals a faint galaxy in the F606W image
(F606W(AB) = 27.0 $\pm 0.3$) and a somewhat brighter galaxy in the F160W 
observations (F160W(AB) =24.5 $\pm 0.2$). 
Interestingly the centroid of the galaxy in the F160W observations
is offset $\sim 0.3$\arcsec\ west of the centroid in the F606W observations.
This indicates either a color gradient within the host, or some 
continuing contribution from the afterglow.  Further F160W observations will be
necessary to distinguish between these possibilities.
Furthermore there are five extremely red objects (EROs) in the field which
are exceptionally faint (and low surface brightness) or 
undetected with ACS, but relatively bright in
F160W. Some of these are exceptionally red ($F606W(AB)-F160W(AB) >
4$), and would fall amongst the
reddest galaxies observed in GOODS with R$-$K$_S > 4.35$ (AB), a
population which comprises only $\sim$ 20\% of the total number of
EROs (Moustakas et al. 2004). Galaxies as red as this have an areal
density on the sky of $0.41 \pm 0.05$ arcmin$^{-2}$ (Gilbank et al. 2004) and
thus in our observed field of only 0.5 arcmin$^{2}$ they are overdense
by a factor of 20, although it should be noted that EROs tend to be
highly clustered. These objects may either represent a highly
redenned population at moderate redshift ($\sim 2$) or
could lie at very high redshift ($z \sim 5$), although in the latter case they would
be significantly brighter than typical Lyman break galaxies at comparable redshift
(e.g. Lehnert \& Bremer 2003).

\section{Discussion}

The afterglow of GRB 060121 exhibits a bright X-ray
afterglow, but is very faint in the optical (R$\sim$23 only 2 hours
after the burst). The optical afterglow is somewhat fainter than previously
reported optical afterglows of short duration bursts (e.g. Hjorth et
al. 2005a; Soderberg et al. 2006). In fact, the observed optical to X-ray
spectral slope is flatter than expected for the fireball
model ($\beta_{\mathrm{OX}}$ = 0.2), rendering GRB 060121 a dark burst as defined by
Jakobsson et al. (2004).
Indeed the direct extrapolation of the X-ray flux into the optical
window using the technique of Rol et al. (2005) also 
places the R-band magnitude significantly fainter than its expected level. However
the value of $\beta_{\mathrm{KX}} = 0.6$ is broadly consistent with the
extrapolation of the X-rays to the K-band assuming that the cooling break
lies between the two frequencies. Indeed the X-ray and optical lightcurves, 
shown in Figure 1 demonstrate a
slower decay in the optical than the X-ray and are also consistent with the
presence of the cooling break between the two frequencies.  These
parameters (with exception of the $\beta_{\mathrm{OX}}$ which is discussed
below) are consistent with those that would
be expected from a $p \sim 2.2$ afterglow model.

The
extrapolation of the optical lightcurve to late times is above the
limit obtained via the {\it HST} observations and indicates
that a temporal break has occurred. There is no break in the X-ray
lightcurve out to 10 days post-burst, however optical monitoring
ceased after 1.5 days. If the break in the R-band is due to the
jet-break then it would have occurred $>$ 10 days after the burst and
the required slope would be $\alpha > 1.7$. For generic ambient medium
parameters at $z=3$ (see below) this would correspond to a beaming
angle of $>7$ degrees. However, if this break were the cooling break
then it could have occurred in the R-band and been unobserved in the
X-ray. If it occurred at the time of the final R-band observations
then the late time slope is only constrained to be $\alpha > 1.1$,
consistent with the X-ray slope, and thus with the motion of the
cooling break. However, the value of $\beta_{\mathrm{KX}} \sim 0.6$ indicates
that the cooling break lies close to the X-ray band at $\sim$ 30
hours, and thus it is unlikely to reach the R-band until very late
times.

A natural explanation of the SED is that the R-band afterglow is
extinguished due to the presence of dust in the host galaxy. This
would be supported by the red R$-$K color and the high column density
measured directly from the X-ray spectrum. The X-ray and (to a lesser
extent) K-band observations are largely unaffected by the presence of
dust, while the R-band (or rest frame B or UV depending on $z$)
exhibits the strongest signature of extinction. The location of GRB
060121 on a possibly edge-on galaxy, in which significant dust
extinction could occur through the disk lends some support to this
theory. Although the GRB does not lie near the centroid of the galaxy
it does apparently lie along the major-axis of the galaxy, 
consistent with an origin in the disk.

The currently most popular model for (most) short-GRBs is that they
originate from compact binary (NS-NS or NS-BH) mergers.  At formation
a neutron star receives a significant natal kick (e.g Arzoumanian et
al. 2002); thus, NS-NS binaries are expected to be kicked from their
birthplace at significant velocities (several hundred
km~s$^{-1}$). Their merger time is governed by the gravitational
radiation timescale and can span a wide range of times from $10^6 - 10^{10}$ years
(e.g. Burgay et al. 2003).  However, even at the shorter end of this
range the NS-NS binary would have travelled $> 100$ pc from its birth
site, and out of the region of star formation in which it formed.
Indeed, with a large kick and long inspiral time, the host galaxy
might then be one of the many other galaxies in the field.
It is unlikely that the faint R-band magnitude of the burst was caused
by its being in the IGM -- both the K-band and X-ray were reasonably
bright.

An alternative explanation for the faint R-band afterglow is that the
burst lies at significantly higher redshift ($z \sim 5$). In this case
the low value of $\beta_{\mathrm{OX}}$ would be explained by the Lyman-$\alpha$
break lying within the R-band, while the K-band would be unaffected
and would lie on the extrapolation of the X-ray afterglow (as is
observed). However, in this case we might expect to observe a stronger
break between the R and I-bands which.  For example GRB 000131 at
$z=4.5$ had R-I = $1.2$ (Andersen et al. 2000) and GRB 050814 at
$z=5.3$ had R-I $\sim$ 2.5 (Jakobsson et al 2006) while the measured
colour of GRB 060121 (extrapolating I to the same epoch as our first
R-band observation) is R-I = 0.90 $\pm 0.36$.  This value has a large
error due to the low signal to noise in our images and in fact is consistent
with the colour of GRB 000131 (although also with a typical $\nu^{-1}$
spectrum). A higher redshift, comparable to that of GRB 050814 is
apparently ruled out, although $z \sim 4.5$ certainly remains
plausible.  Further observations of the host galaxy,
especially in the I-band will be helpful in distinguishing between a
strong break due to Lyman$-\alpha$ and gradual reddening.

Finally it is also important to consider whether GRB 060121 could in
fact be due to a collapsar. Although its spectrum is moderately hard
the duration of $\sim 2$~s puts it within a region of the BATSE
hardness-duration plot that contains an admixture of long, short, and
possibly intermediate duration bursts (Horvath et al. 2005).  Indeed,
some classical long duration bursts have rest frame durations of
$<$2~s (e.g. GRB 000301C Jensen et al. 2000; GRB/XRF 050416 Sakamoto
et al. 2006; GRB 060206 Fynbo et al. 2006).  However, if at $z\sim3$,
then the rest-frame duration of GRB 060121 becomes $<0.7$~s, while
at lower redshift its host is significantly redder than the typical colors
of long burst hosts (Christensen et al. 2004). Making a collapsar origin less plausible in
this case.

\section{Conclusions}

We have presented deep optical and X-ray observations of the
short-hard GRB 060121. These observations demonstrate that both the
afterglow and host galaxy of GRB 060121 are significantly fainter and
redder than previously optically-detected short bursts.
 
We believe the most likely explanation of its properties is that GRB
060121 lies at higher redshift than previously observed short bursts,
$z > 2$, and probably is furthermore extinguished. If this is a compact
binary merger then significant dust extinction is surprising since it
would be expected that the binaries should receive significant natal
kicks. Thus in this case either the binary received very little kick,
or (by chance) its line of sight passed through an intervening galaxy.

Interestingly, GRB 060121 lies rather closer, in the hardness-duration
sense, to the typical short-burst as seen by BATSE than the other
well-localised short bursts studied to date.  GRBs 050709 and 050724,
in particular, were all very much at the spectrally soft end of that
distribution (Hjorth et al. 2005a).  This raises the question as to whether a significant
proportion of the bursts classified by BATSE as short could in fact be
a much higher redshift population than those seen recently by \swift\
and \hete. If so, then the width of the luminosity
function of short bursts would be much wider than has previously been 
considered (e.g. Piran \& Guetta 2005), 
with consequent implications for progenitor models.

\section*{Acknowledgements}
We thank Jochen Heidt and Caroline Villforth for assistance 
with our NOT observations. 
AJL, NRT \& ER are supported by PPARC. The Dark
Cosmology Centre is funded by the Danish National Research
Foundation. Based in part on observations made with the NASA/ESA
 {\it Hubble
Space Telescope,} obtained at the Space Telescope Science Institute,
operated by the Association of Universities for Research in
Astronomy, Inc., under NASA contract NAS 5-26555. 
These observations
are associated with program 10870. This research was supported in
part by the National Science Foundation under Grant No. PHY99-07949.

\end{document}